\documentclass[12pt]{article}

\usepackage{setspace}
\usepackage{amsmath,amssymb,amscd}
\usepackage{amsfonts}
\usepackage{color}

 \newcommand{\bea}{\begin{eqnarray}}
\newcommand{\eea}{\end{eqnarray}}
\newcommand{\be}{\begin{equation}}
\newcommand{\ee}{\end{equation}}
\newcommand{\ba}{\begin{align}}
\newcommand{\ea}{\end{align}}

\newcommand{\ZZ}{\mathbb{Z}} % Interi
\newcommand{\Tr}{{\rm {Tr}}}

\newcommand{\cH}{\mathcal{H}}

\newcommand\rref[1]{(\ref{#1})}

\newcommand{\ie}{{\it i.e.~}}

\newcommand{\ls}{\ell_s}

\newcommand{\tD}{{\tilde{\Delta}}}

\begin{document}

\title{\bf{String Universality for Permutation Orbifolds}}

\author{Alexandre Belin$^\bigstar$, Christoph A. Keller$^\spadesuit$ and Alexander Maloney$^{\bigstar}$}

\maketitle
\begin{center}
$^\bigstar$\ {\it  Department of Physics, McGill University, Montr\'eal, Canada } \\ \bigskip
$^{\spadesuit}$\ {\it NHETC, Rutgers, The State University of New Jersey, Piscataway, USA}\\ \smallskip

\vspace{2em} 
\texttt{alexandre.belin@mail.mcgill.ca, keller@physics.rutgers.edu, maloney@physics.mcgill.ca}
\end{center}
\vspace{3em}

\begin{center}
{\bf Abstract}
\end{center}

The hypothesis that every theory of quantum gravity in AdS$_3$ is a dimensional reduction of string/M-theory leads to a natural conjecture for the density of states of two dimensional CFTs with a large central charge limit.
We prove this conjecture for 2D CFTs which are orbifolds by permutation groups.
In particular, we characterize those permutation groups which give CFTs with well-defined large $N$ limits and can thus serve as holographic duals to bulk gravity theories in AdS$_3$.
We then show that the holographic dual of a permutation orbifold will have a Hagedorn spectrum in the large $N$ limit.
This is evidence that, within this landscape, every theory of quantum gravity with a semi-classical limit is a string theory.

\newpage

\section{Introduction and Summary}

AdS/CFT \cite{Maldacena:1997re, Witten:1998qj, Gubser:1998bc} is an important tool in the study of quantum gravity, but has not answered perhaps the most fundamental question: which classical theories of gravity can be quantized?  
Known quantum theories of gravity with a semi-classical limit appear to contain many degrees of freedom in addition to the metric, such as those coming from string theory.  
But we have not found a precise characterization of the constraints which must be placed on the semi-classical spectrum in order for the theory to have a quantum mechanical completion.
The goal of the present paper is to outline a strategy to address this question using the AdS/CFT correspondence, and to implement this strategy in a simple setting: theories of quantum gravity dual to permutation orbifold CFTs in two dimensions.

Our starting point is the observation that every conformal field theory can be interpreted as a fully consistent theory of quantum gravity in asymptotically Anti-de Sitter Space, and that conversely every theory of AdS gravity will, in principle, define a set of CFT correlation functions. 
CFT operators are dual to states in the Hilbert space of the bulk gravity theory, and CFT correlation functions for operators of sufficiently low dimension define asymptotic scattering amplitudes for a gravity theory in AdS.  Of course, the gravity theory dual to a typical CFT will in general be strongly coupled, so  the bulk gravitational description is not particularly useful.  
We are really interested in theories of gravity which have a semi-classical limit.  
We therefore consider not a single CFT, but rather a family of CFTs labelled by a parameter $N$ which can be interpreted the AdS radius in Planck units. 
We then seek to infer general properties about the bulk theory by considering the space of conformal field theories in the large $N$ limit, where the gravity theory becomes weakly coupled.

We should emphasize that, in the present context,  by a ``weakly-coupled" theory of gravity we do not necessarily mean just perturbative Einstein gravity coupled to matter.  We only mean that as $N\to\infty$ the Planck length is small in AdS units, so that gravitational backreaction is negligible.
We do not demand that the bulk theory is local on the AdS scale; this would require additional constraints, such as the existence of a large gap in the CFT spectrum \cite{Heemskerk:2009pn}. 
We would, for example, be happy to consider theories of gravity which have as their $N\to\infty$ limit a classical ($g_s\to0$) string theory with string length of order the AdS scale.  Such theories are expected to be dual to weakly coupled gauge theories in the large $N$ limit.

\subsection{The Semi-Classical Limit of AdS$_3$/CFT$_2$}

We will focus on the case of AdS$_3$/CFT$_2$.  Although the strategy described below generalizes to higher dimensions, we will be able to make exact statements in three dimensions.  In this case it is convenient to consider a family of CFTs labelled by their central charge $c$.  In the bulk, this central charge is interpreted as the AdS radius in Planck units \cite{Brown:1986nw}
\be
c= {3 \ell \over 2 G}
\ee  
where $\ell$ is the AdS radius and $G$ is Newton's constant.
We will denote by $\rho( h, {\bar h})$ the number of CFT states with left and right-moving dimension $(h, {\bar h})$.  
The conformal dimensions $(h, {\bar h})$ are related to the energy and angular momentum of the states in the bulk by 
\be
E={h +{\bar h} \over \ell} = \frac{\Delta}{\ell},~~~~~J=h - {\bar h} \in \ZZ
~\ee  
where $\Delta = h + {\bar h}$ is the total scaling dimension.  Our goal is to characterize the density of states of the theory in the limit of large $c$.

Before considering specific families of CFTs, let us ask what densities of states we might expect for different theories of bulk gravity. The behaviour of the density of states $\rho(E)$ will depend on the energy $E$.\footnote{We  omit the dependence on $J$ for simplicity, imagining that $J$ remains fixed and of order one as $E$ is varied.}
At very high $E$, much bigger than  the three dimensional
Planck mass $M_{Pl}= {8 G}^{-1} $, 
 the typical state will be a BTZ black hole and the density of states is given by the Bekenstein-Hawking entropy \cite{Banados:1992wn}:
\begin{equation}
\label{BTZ}
{\rm Black~Holes:}~~~~~~~~~\log \rho(E) \sim \left(E \over E_0\right)^{1/2}
\ ,
\end{equation}
with $E_0 = {G \over 2 \pi^2 \ell^2}$.
On the CFT side, this behavior is well understood: the asymptotic density of states is given by the Cardy formula \cite{Cardy:1986ie, Strominger:1997eq}, whose exact
range of validity was  analyzed in \cite{Hartman:2014oaa}.

In this paper we are interested in the behaviour below the Planck scale,  
where the density of states is not given by the black hole entropy.
In particular, we wish to characterize the perturbative degrees of freedom which are present in our theory of gravity at  weak coupling.  This information is contained in $\rho(E)$ with $E\lesssim M_{Pl}$.   

To begin, let us ask what density of states would be characteristic of Einstein gravity plus matter in  AdS$_3$.  When $E$ is below
the mass $m$ of the lightest particle of the bulk theory (assuming $m>0$) the theory 
is described by general relativity in AdS$_3$.  
This theory has no local degrees of freedom, but it does possess perturbative states, known as boundary gravitons, which come from non-trivial diffeomorphisms applied to the AdS ground state \cite{Brown:1986nw, Witten:2007kt, Maloney:2007ud}. 
 Counting the degeneracy of these states, we find that when $E$ is large (but still less than $m$ or $M_{Pl}$):
\begin{equation}
\label{pure}
{\rm Pure~ Gravity:}~~~~~~~~~\log \rho(E) \sim \left(E \over E_0\right)^{1/2}
\ 
\end{equation}
with $E_0 = {3 \over 4 \pi^2 \ell}$. 
It is important to note that even though
the scaling is the same as  (\ref{BTZ}), the value of $E_0$ here is much smaller.
In particular, (\ref{pure}) has the same form as the Cardy formula because the boundary excitations of a topological theory in AdS are essentially localized at the two dimensional boundary, and thus exhibit the characteristic growth of states of a two dimensional field theory.\footnote{The states being counted in here are the Virasoro descendants of the identity, whose high energy density of states is given by the Cardy formula with $c=1$.  For related comments, see \cite{Castro:2010ce, Castro:2011ui}.}

If there are no other dynamical degrees of freedom, then
(\ref{pure}) is valid all the way up to the Planck scale $E\lesssim M_{Pl}$.
This is the characteristic behaviour of the extremal CFT partition functions conjectured by Witten to be dual to pure gravity \cite{Witten:2007kt}.
In general, the growth (\ref{pure}) would be expected of any topological theory of gravity in AdS${}_3$ -- such as Einstein-Chern-Simons theory or a pure higher-spin theory of gravity based on  $SL(N)$ Chern-Simons theory, with $N$ held fixed as $c\to\infty$ (as in \cite{Henneaux:2010xg,Campoleoni:2010zq}).  The precise value of $E_0$ will differ in these examples.  
A family of CFTs with a large $c$ limit whose spectrum obeyed (\ref{pure}) all the way up to the Planck scale 
would be a natural candidate for a ``pure" theory of quantum gravity in AdS, i.e. a theory of gravity with only metric degrees of freedom. No examples of large $c$ CFTs with this behaviour are  known.  Extremal CFTs with large central charge may not exist \cite{Gaberdiel:2007ve, Gaberdiel:2008pr, Gaiotto:2008jt}.

If the bulk theory has a dynamical local field of mass $m$, then above this scale
the density of states will be that of a local field theory in AdS. 
In particular, for $E$ much larger than $m$ (but still  less than $M_{Pl}$) 
the density of states
will take the usual form for a three dimensional  field theory at finite volume:   
\begin{equation}
\label{local}
{\rm Local~ QFT_3:}~~~~~~~~~\log \rho(E) \sim \left(E \over E_0\right)^{2/3}
\ .
\end{equation}
where $E_0$ is set by the AdS length scale.
A family of CFTs with a large $c$ limit whose density of states obeys (\ref{local}) all the way up to the Planck mass $M_{Pl}$ would be a candidate dual to a local quantum field theory of gravity, i.e. a theory of a  metric coupled to a finite number of local degrees of freedom. 

Slightly more generally,
we could consider local QFT in $d>3$ dimensions (such as 10 or 11 dimensional supergravity) which is compactified down to AdS$_3$ on a manifold of finite volume.  At energies above the Kaluza-Klein scale $m_{KK}$, the extra dimensions will become visible
and the scaling will be:  
\be
\label{HigherD}
{\rm Local~QFT}_{d}: ~~~~~~~~~\log \rho(E) \sim \left(E \over E_0\right)^{d-1\over d}
~
\ee
where $E_0$ depends on both the AdS$_3$ radius and the compactification volume.

All known examples of AdS/CFT have a density of states which increases more rapidly than a local quantum field theory in three dimensions. 
This reflects the fact that the bulk theories describe gravity coupled to an infinite number of degrees of freedom, with a density of states that increases rapidly at high energy.
Let us consider string theory at weak coupling $g_s \ll 1$.
The characteristic feature of this theory is an exponential density of states above the string scale:
\be
\label{hagedorn}
{\rm String~Theory:}~~~~~~~~~\log \rho(E) \sim {E \over E_0}~~~~~
~ 
\ee
where $E_0\sim \ls^{-1}$ is set by the string scale $\ls$.
The density of states (\ref{hagedorn}) would, if it were valid at arbitrarily high energy, lead to a divergence in the free energy at the Hagedorn temperature $T_H=E_0$. For interacting string theories ($g_s\ne 0$)
the Hagedorn spectrum will be cut off at the Planck scale, where the strings will collapse
into black holes.  Thus a CFT dual to a string theory will, at large central charge, have a density of states given by (\ref{hagedorn}) for $E\lesssim M_{Pl}$ and by (\ref{BTZ}) for $E \gg M_{Pl}$.

We will regard any family of CFTs with a large $c$ limit which obeys (\ref{hagedorn}) for $E\lesssim M_{Pl}$ as a string theory, in the sense that the perturbative density of states matches that of a theory of extended objects. 
This notion of a ``string theory" is quite general, and does not necessarily guarantee that the bulk theory has a simple worldsheet description or obeys other locality or analyticity properties typically found in worldsheet string theories.

Other types of growth are possible.  
In particular, a compactification of $M$ theory down to three dimensions would presumably have only local supergravity states up to the Planck scale.  Thus in the semi-classical limit we would have
\be
\label{Mtheory}
{\rm M-Theory}: ~~~~~~~~~~\log \rho(E) \sim \left(E \over E_0\right)^{10\over 11}
~.
\ee

It is natural to conjecture that every quantum theory of gravity with a semi-classical limit is a string theory or a compactification of M-theory.
In view of the above, this means that the dual CFT should have 
a regime where the growth of states is of the form (\ref{hagedorn})
or (\ref{Mtheory}).  We therefore make the following
\begin{itemize}
\item[]
\noindent{\bf Conjecture} (String Universality): Every family of CFT$_2$'s with a well-defined large central charge limit will have
a regime where
\be\label{stringuniversality}
\log \rho(\Delta) \gtrsim \left(\Delta \over \Delta_0\right)^{10\over11}
~
\ee
for some value of $\Delta_0$.
\end{itemize}
\noindent
As we  discuss below, we will need to define precisely what we mean by ``well-defined large central charge limit."
We note that the spectrum will only have the rapid growth (\ref{stringuniversality}) below the Planck scale, i.e. for CFT operators with scaling dimension $\Delta \lesssim c$.  At high energies ($\Delta \gg c$) the spectrum will have the slower Cardy growth (\ref{BTZ}) associated with black hole entropy. 

At this point we know of no counter-example to the string universality conjecture.\footnote{An interesting set of potential examples are minimal model CFTs.  Certain minimal models were argued to be dual to ``pure" theories of quantum gravity in \cite{Castro:2011zq}.  However, this set of theories does not have a semi-classical (large central charge) limit.  A family of ${\cal W}_N$ minimal models with a large central charge limit was discussed in \cite{Gaberdiel:2010pz}, although it is not clear
that by themselves these lead to a consistent bulk gravity theory unless
embedded in string theory \cite{Gaberdiel:2014cha}.  
In any case, even though  the density of descendant states  
grows as in local QFT$_3$ \cite{Castro:2010ce}, in the semi-classical limit the number of light primary fields in these theories has
super-Hagedorn growth \cite{Gaberdiel:2013cca}. }
However, a general proof of the conjecture may require a full classification of two dimensional CFTs at large central charge.  This is a difficult task.  In this paper we will  be content to prove that the conjecture holds for a certain broad class of large $c$ CFTs.  

We now describe our general approach.  We will consider a family of CFTs labelled by an integer $N$.
We will assume that each CFT is unitary and has a discrete spectrum $\rho_N(h, {\bar h})$ with a normalizable ground state.
The central charge will be proportional to
$N$, so the bulk gravity theory is weakly coupled when $N\to\infty$.  However, not every family of CFTs will lead to a well-defined semi-classical bulk theory.
We must demand that the density of states is well behaved as we take $N\to\infty$, in the sense that
the limit 
\be
\label{rhoinfty}
\rho_\infty (h, {\bar h}) = \lim_{N\to \infty} \rho_N(h, {\bar h})
~\ee
exists and is finite.
The function $\rho_\infty(h, {\bar h})$ is the density of states of the bulk theory in the limit where gravity becomes weakly coupled. 
Note that, because we take $N\to\infty$ while holding $h, {\bar h}$ fixed, black hole states are not counted in (\ref{rhoinfty}). So the asymptotics of the density of states $\rho_\infty(h, {\bar h})$ contains information about the short-distance physics of the weakly coupled  bulk theory, rather than long-distance (black hole) physics.\footnote{In particular, the asymptotics of $\rho_\infty(h, {\bar h})$ are {not} determined by the Cardy formula. The partition function at finite $N$ is modular invariant, but the partition function built out of $\rho_\infty(h, {\bar h})$ is not.}
In the present paper we will not demand that an actual limit CFT exists by, for example, demanding that correlation functions or higher genus partition functions converge as $N\to\infty$.  We will simply demand that $\rho_\infty(h, {\bar h})$ is finite, and leave a more refined analysis to later work.
This simple constraint alone will lead to remarkable results.

For most families of CFTs the limit
(\ref{rhoinfty}) will not exist.
Our strategy is to understand what constraints the finiteness of $\rho_\infty$ places on our family of CFTs.
For example, the theory of N free bosons -- or any theory built out of N non-interacting copies of a starting ``seed" CFT -- will have a low lying spectrum which diverges badly as $N\to\infty$.   
To obtain a finite spectrum, one needs to consistently project out all but a finite number of states in the $N\to\infty$ limit.
This is essentially the reason why AdS theories of gravity in higher dimensions are dual to gauge theories: to obtain a finite spectrum one needs to impose the singlet constraint under some global symmetry, and gauge theories are the only way to enforce such a singlet constraint (via the Gauss law constraint) using local dynamics.  In two dimensions, the projection onto the singlet sector of a global symmetry leads to an orbifold CFT.  
These theories are the subject of this paper.

\subsection{The Landscape of Permutation Orbifolds}

The simplest way to construct a CFT with large central charge is to take $N$ non-interacting copies of a particular seed CFT ${\cal C}$.  If ${\cal C}$ has central charge $c$, then the resulting tensor product CFT ${\cal C}^{\otimes N}$ will have central charge $N c$.  As $N\to\infty$, the density of states $\rho_\infty(h, {\bar h})$ of ${\cal C}^{\otimes N}$  will diverge. 
In order to project out states, we must use the fact that the product CFT ${\cal C}^{\otimes N}$ has a large global symmetry.
It is invariant under the $S_N$ permutations of the individual copies of ${\cal C}$ in ${\cal C}^{\otimes N}$.
So we can consider the orbifold of ${\cal C}^{\otimes N}$ by any permutation
group $G_N \subseteq S_N$:
\be
\mathcal{C}_{G_N}=\mathcal{C}^{\otimes N}/G_N \ \ \ \ \ \ \ \ \ \ \  G_N \subseteq S_N\ .
\ee
This provides a natural landscape of large $N$ CFTs.  Essentially, we have one theory of quantum gravity for each choice of permutation group $G_N\subseteq S_N$.

These permutation orbifolds are the two dimensional analogues of free theories in
higher dimensions, and $G_N$ is analogous to a choice of gauge group.\footnote{This analogy can be made sharper in special cases, such as when ${\cal C}$ is a free boson.  One could then consider orbifolds by larger discrete subgroups or even by continuous groups such as $O(N)$ or $U(N)$ (as in e.g. \cite{Gaberdiel:2011aa, Banerjee:2012aj}). However, the permutation group $S_N$ will be the global symmetry group of ${\cal C}^{\otimes N}$ for a typical seed theory, so we expect that the class of orbifolds studied in this paper are generic.} The different copies of $\mathcal{C}$ in
$\mathcal{C}^{\otimes N}$ do not interact with each other,
and orbifolding  does not fundamentally change this fact. Indeed, correlation functions
of twisted sector states satisfy selection rules
similar to free theories. It is therefore natural to expect that their bulk duals are stringy, with string length of order the AdS scale.
This is indeed what we will find. However, we will see that permutation orbifolds have a much richer structure than one might expect from free gauge theory. 
For example, many of the properties of $\mathcal{C}_{G_N}$ depend on the choice of the permutation group $G_N$ but are largely independent of the choice of seed CFT ${\cal C}$.  In particular, the large $N$ properties of the spectrum we are interested in will depend only on the central charge $c$ of the seed CFT and a choice of a family of permutation groups.

We will prove the string universality conjecture (\ref{stringuniversality})
for the landscape of permutation orbifolds.
The intuition behind the proof is easy to understand.
If $G_N$ is too small, then we do not project out enough states, 
and $\rho_\infty(h,\bar{h})$ will diverge for some finite values of $h, {\bar h}$. 
Thus $G_N$ must be a rather large subgroup of $S_N$.
On the other hand,
an orbifold does not merely project out states,
it also introduces new twisted sector states.  In the gauge theory language, these are states with non-trivial holonomy around the spatial circle.
We will show that $G_N$ must be so large that the twisted sector states lead to a Hagedorn density.  For the symmetric product orbifold ($G_N=S_N$) this is a well known result; it is the ``long string" behaviour of symmetric products (see e.g. \cite{Vafa:1995bm, Strominger:1996sh, Dijkgraaf:1996xw, Maldacena:1998uz, Seiberg:1999xz}).  Our result shows that this long string picture persists for any subgroup of $S_N$ that leads to a well defined bulk gravity theory.

In particular, we will show that for any family of permutation orbifolds with a well-defined large $N$ limit there is a regime where
\be
\label{symm}
\log \rho_{N} (E)\sim {E \over E_0},~~~~~~~~~E_0 = {1 \over 2 \pi \ell}
~.\ee
The Hagedorn temperature, and hence the string-scale associated with this growth of states, is the AdS scale. 
This is just as in the $S_N$ case. However, we find an important distinction between the symmetric product orbifold $S_N$ and a more generic permutation orbifold $G_N$.  For symmetric products, equation (\ref{symm}) holds for any $E\gg \ell^{-1}$, i.e. for states which are heavy in AdS units but still far below the Planck scale.  For generic permutation orbifolds, we are only able to show that (\ref{symm}) holds for states that are close to the Planck scale, $E\lesssim M_{Pl}$.  So in principle there could be another energy scale $E_1$, with $\ell^{-1} \ll E_1 \lesssim M_{Pl}$ such that (\ref{symm}) holds only above $E_1$.\footnote{
The results of \cite{Hartman:2014oaa} give an upper 
bound $E_1 < 2M_{Pl}$, since if the spectrum is sub-Hagedorn all the way up to to the Planck scale
it has a ``sparse light spectrum,"
from which it  follows that
$\rho(E=2M_{Pl}) \sim e^{2\pi E/\ell}$.
}
This is not the typical situation in perturbative string theory, where a single dimensionful scale -- the string length --  sets both the Hagedorn temperature and the value of the energy at which the spectrum becomes Hagedorn.  

However, we are still able to strongly constrain the spectrum of general permutation orbifolds far below the Planck scale.
In particular, in section~2 we will show that for permutation orbifolds
\be\label{nonlocal}
\rho_\infty(E) \gtrsim \exp \left(\frac{c\pi^2 E \ell/3}{\log c\pi^2 E\ell/3} \right)
\ee
when $E\gg {c \ell}^{-1}$. Here $c$ is the central charge of the seed theory $\mathcal{C}$, so is independent of $N$.  As this exceeds the growth (\ref{HigherD}), this means that the bulk dual is non-local on the AdS scale.  Thus the semi-classical limit of any theory of quantum gravity in our landscape cannot be a local quantum field theory.  We note that the scale at which the bulk theory is non-local is determined by the central charge $c$ of the seed CFT, so this parameter is playing the role similar to the 't Hooft coupling in higher dimensional AdS/CFT.
In section~3 we give evidence that (\ref{nonlocal}) is probably an underestimate, and that the growth is actually Hagedorn even for energies far below the Planck scale.  Thus we conjecture that the parameter $E_1$ described above is always of order $\ell^{-1}$, although we have not been able to prove it. 

\vspace{.5cm}\noindent
\emph{Note added:} While we were in the process of
writing up this note, the paper \cite{Haehl:2014yla} appeared,
which partly overlaps with the results presented here.

\section{The Untwisted Sector Partition Function}

In constructing our orbifold CFTs, we start with a ``seed" CFT ${\cal C}$ with 
central charges $c$ and $\bar{c}$.  The  partition function is
\be
Z(\tau)=\Tr_{\cH} q^{L_0}\bar{q}^{\bar{L}_0}
\ee
where $q=e^{2\pi i \tau}$ with $\tau = i\beta/2\pi+\mu$,
where $\mu$ is the angular potential and $\beta$
is the inverse temperature. The parameter $\tau$ can be interpreted as the modular parameter of the torus. 
Note that we have defined the partition function without
the customary shift in the vacuum energy, so that the ground state has $h={\bar h} = 0$.
Written in terms of the density of states, we have
\be
Z(\tau) = \sum_{h,\bar{h}} \rho(h,\bar h) q^{h} \bar{q}^{\bar h}\ ,
\ee
or in the case of no angular potential
\be
Z(\beta) = \sum_\Delta \rho(\Delta)e^{-\beta\Delta}\ .
\ee

We will consider a family of CFTs $\mathcal{C}_{G_N}$ defined by taking the $N$ copies of ${\cal C}$ and orbifolding by the action of a permutation group $G_N$:
\be
\mathcal{C}_{G_N}=\mathcal{C}^{\otimes N}/G_N \ \ \ \ \ \ \ \ \ \ \  G_N \subseteq S_N
~.
\ee
The partition function  of $\mathcal{C}_{G_N}$ will be written as 
\be
Z_{G_N}(\tau)=\sum_{h,\bar{h}} \rho_{N}(h,\bar{h}) q^{h}\bar{q}^{\bar h} \,.
\ee
In order for the dual gravity theory to be well defined as $N\to \infty$, the limit 
\be
\rho_\infty(h,\bar{h})=\lim_{N\to\infty} \rho_N(h,\bar{h})<\infty
\ee
must exist and be finite.
In this section we will analyze the properties of $\rho_\infty$ and understand the constraints that the finiteness of $\rho_\infty$ places on $G_N$. 
The case of the symmetric product orbifold $G_N=S_N$ has been well studied \cite{Dijkgraaf:1996xw,Bantay:2000eq} and leads to a finite $\rho_\infty$.  Another simple example is $G_N=\ZZ_N$, which was studied in \cite{Klemm:1990df}.   In this case, simple combinatorics shows that $\rho_\infty$ will diverge. 
Thus in a sense we are looking for groups which are larger than $\ZZ_N$, but not necessarily as big as the full symmetric group $S_N$.

The orbifold theory
$\mathcal{C}_{G_N}$ has two types of states: twisted and untwisted.  
The untwisted states are found by taking the states in the $N$-fold tensor product of the original CFT and projecting onto the states which are invariant under the action of $G_N$.
The twisted  states come from changing the boundary conditions of the fields by the action of an element of $G_N$ as we go around the spatial circle.
The full partition function is the sum of the untwisted and twisted sector partition functions:
\be
Z_{G_N}(\tau)=Z_{G_N}^u(\tau)+Z_{G_N}^{t}(\tau)
~.\ee
In this section we will discuss the untwisted sector, and use the finiteness of $\rho_\infty$ in the untwisted sector to constrain $G_N$.  We will discuss twisted sector states in section~3.

\subsection{Untwisted Sector States}

The Hilbert space of the untwisted sector is $\cH^u_{G_N} = (\cH\otimes \cdots \otimes \cH) / G_N$ where $\cH$ is the Hilbert space of the seed theory.  We will compute the resulting untwisted-sector partition function
\be
Z^u_{G_N}(\tau) = \Tr_{\cH^u_{G_N}} q^{h}\bar{q}^{\bar{h}} = \sum_{\Delta} \rho^u_N(h,\bar{h}) q^h\bar{q}^{\bar h}
\ee
in terms of the partition function $Z(\tau)$ of the seed theory.
For the symmetric group $S_N$, we can write explicit expressions
for $Z^u_{S_N}$.

We start by noting that, because $G_N$ is a subgroup of the symmetric group, any element $g\in G_N$ can be written in a unique way as a product of disjoint cycles of length $i=1,\dots, N$. With $j_i(g)$ the number of cycles of length $i$, we have schematically 
\be\label{cycledecomp}
g \sim ({\bf 1})^{j_1(g)}\dots({\bf N})^{j_N(g)} \ ,
\ee
where $({\bf i})$ represents a cycle of length $i$.
We have
\be
N = \sum_{i=1}^N i j_i(g) \,,
\ee
so the vector ${\bf j}(g) = (j_1,\dots, j_N)$ describes a partition  of the integer $N$ and can be represented by a Young diagram with $j_i$ rows of length $i$. 
If $G_N$ is the symmetric group $S_N$, then the cycle decomposition (\ref{cycledecomp})
labels the conjugacy classes of $G_N$. For
a general permutation group this is not the case: multiple distinct conjugacy classes of $G_N$ might have the same cycle decomposition.
But, as we will see, ${\bf j}(g)$ still contains the information we need.

  The cycle index of $G_N$ is defined to be the following polynomial in $n$ variables: 
\be
\chi(G_N; s_1,\dots,s_N) = {1\over |G_N|} \sum_{g\in G_N}s_1^{j_1(g)}\dots s_N^{j_N(g)} 
~.\ee
Let us denote by $A_{\bf j}$ the number of elements of $G_N$ which have an expression in terms of disjoint cycles of the form $({\bf 1}) ^{j_1}\dots({\bf N})^{j_N}$.  Then the cycle index is the generating function for $A_{\bf j}$:
\be
\chi(G_N; s_1,\dots,s_N) = {1\over |G_N|} \sum_{\bf j}A_{\bf j} s_1^{j_1}\dots s_N^{j_N} 
\ee
where the sum is over all partitions $\bf j$ of $N$.

To count the number of states in $\cH^u_{G_N}$ with given conformal dimensions $h$ and $\bar{h}$ 
we project onto the invariant states by summing over all $g\in G_N$ inserted in the partition function:
\be\label{projectuntwisted}
Z^u_{G_N}(\tau) = \frac{1}{|G_N|}\sum_{g\in G_N}\Tr_{\cH^{\otimes N}} q^{L_0}\bar{q}^{\bar{L}_0}g\ .
\ee
Note that even if $G_N$ is not the full symmetric group,
the underlying ${\cal C}^{\otimes N}$ theory still has global $S_N$ symmetry. 
This means that the trace in (\ref{projectuntwisted}) only
depends on the conjugacy class of $g$ in $S_N$,
which is given by its cycle decomposition (\ref{cycledecomp}).
We can thus apply Polya's enumeration theorem to compute $Z_{G_N}^u$ in terms of the partition function $Z(\tau)$ of the original seed theory.  This gives
\bea\label{Polya}
Z^u_{G_N}(\tau) &=& \chi(G_N; Z(\tau),\dots, Z(n\tau)) \cr
&=& {1\over |G_N|}\sum_{\bf j}A_{\bf j} Z(\tau)^{j_1}\dots Z(N\tau)^{j_N}\ .
\eea
We will introduce the following notation. For any function over $G_N$, say $O(g)$ with $g \in G_N$, we will define the group average 
\be
\langle O \rangle= {1\over |G_N|} \sum_{g} O(g)
\ee
which has been normalized so that $\langle 1 \rangle  =1$.  Then 
\bea\label{Polya2}
Z^u_{G_N}(\tau) &=&
\langle Z(\tau)^{j_1}\dots Z(N\tau)^{j_N}\rangle \ .
\eea
We see that the untwisted sector partition function is determined by the statistical distribution of the cycle decompositions ${\bf j}(g)$ of elements of $G_N$.

\subsection{Properties of $G_N$}

We now want to discuss what properties $G_N$ must
satisfy in order for the limit $\rho^u_\infty = \lim_{N\to\infty} \rho^u_N $ to
exist and be finite.  From (\ref{Polya2}), we see that this will constrain the statistics of the cycle decomposition ${\bf j}(g)$ of a typical element in $G_N$.

For the rest of this section, we will for convenience take the seed theory to be a holomorphic CFT with central charge $c=24$. This will greatly simplify the notation and will not change the main results that follow.  The partition function of the seed CFT is then
\be
Z =1+ \rho(1)q+ \rho(2) q^2 + \dots
\ee
where the $\rho(i)$ are strictly positive integers.
Expanding (\ref{Polya}) we get
\bea\label{Zunt}
Z^u_{G_N}(\tau)
&=& \left\langle\prod_{i=1}^N \left(1+ \rho(1) q^i + \rho(2) q^{2i}+\dots \right)^{j_i}\right\rangle 
\cr
&=& \left \langle 1+ \rho(1) \sum_i j_i q^{i} + \dots \right\rangle 
\eea
From the condition that the $\rho_N(i)$ are bounded in the large $N$
limit, we obtain conditions on the number of cycles.
For instance, we have   
\be
\rho^u_{N}(i) = \rho(1) \langle j_i\rangle + \dots
\ee
where $\dots$ are positive numbers.  
Let us define $m_N(i)=\langle j_i \rangle $ to be the average number of times a cycle of length $i$ appears in an element of $G_N$. 
From
\be\label{mconst}
m_N(i) \equiv \langle j_i\rangle \leq \frac{\rho^u_{N}(i)}{\rho(1) }\  < \infty
\ee 
we see that the finiteness of $\rho^u_\infty(i)$ thus implies that
$m_N(i)$ is bounded in the large $N$ limit.
As a point of comparison, we note that $S_N$  has $m_N(i) = 1/i$ for all $N$.  For a more general $G_N$, the cycle decomposition of a typical element may be different.
However, (\ref{mconst})  places an upper bound on the frequency with which cycles of length $i$ can appear in elements of $G_N$.

We can use this to argue that $G_N$ will necessarily have cycles of unbounded length in the large $N$ limit.  This will be very important in section~3, when we discuss the long string picture for twisted states.  In particular, note that
\be\label{cyclecondition}
\sum_{i=1}^N i~ m_N(i) = \sum_{i=1}^N i \langle j_i \rangle =  \left\langle \sum_{i=1}^N i j_i \right\rangle = N 
\ee
From this it follows that $G_N$ has long cycles in the large $N$ limit.  To see this, imagine that $G_N$  had only cycles of length $i\le k$ for some $k$ which is fixed in the large $N$ limit. 
Then the condition $\sum_{i=1}^k i m_N(i)=N$ would contradict our previous result that the $m_N(i)$
are bounded in the large $N$ limit.

In fact, the $m_N(i)$ have a very natural physical interpretation: they constrain the density of states of the untwisted sector.  The free energy of the untwisted sector is
\bea
\label{fbound}
F_{G_N}^u(\beta)  \equiv \log Z_{G_N}^u(\beta) &=& \log \left\langle Z(q)^{j_1}\dots Z(q^N)^{j_N} \right\rangle  
\cr
&\ge& 
\left\langle \log \left(Z(q)^{j_1}\dots Z(q^N)^{j_N} \right)\right\rangle  
\cr
&=& \sum_{i=1}^N m_N(i)  \log Z(i\beta)
\eea
In the second line we have used Jensen's inequality. Thus the cycle distribution of elements of $G_N$ bounds the free energy.
At low temperature we get
\be
\log Z_{G_N}^u(\beta) \ge \rho(1) \sum_{i=1}^N m_N(i) e^{-i\beta} + \dots
\ee
Thus the leading corrections to the 
free energy at low temperature are the moments $\sum_{i=1}^N m_N(i) e^{-i\beta}$ of the $m_N(i)$.

More generally, we can constrain not just the one-point functions $m_N(i)=\langle j_i \rangle$, but also the correlation functions of the $j_i$'s.  
Consider the correlation function $\langle  j_1^{n_1}\cdots j_k^{n_k} \rangle$ for some choice of positive integers $(n_1,\dots,n_k)$ with $n_1+\dots+n_k = M \le N$.
From the $q-$expansion of 
the untwisted partition function (\ref{Zunt}) it is straightforward to check that the untwisted density of states contains a term of the form 
\be
\rho^u_N(M) = \rho(1)^{M} \langle  j_1^{n_1}\cdots j_k^{n_k} \rangle + \dots
\ee
where $\dots$ denotes contributions which are bounded in the large $N$ limit. 
This implies that the correlation functions
\be
m_N(n_1,\dots, n_k) \equiv \langle  j_1^{n_1}\cdots j_k^{n_k} \rangle ~ < \infty
\ee
must be finite in the large $N$ limit for any choice of $(n_1,\dots,n_k)$. Roughly speaking, this means that the cycle distribution of elements of $G_N$ cannot be sharply peaked, nor can it contain any sharp correlations between cycle lengths, in the large $N$ limit.

As an application of this formalism, let us bound the size of the permutation group $|G_N|$.
From the $q$-expansion of (\ref{Zunt}) we find a contribution to $\rho^u_N(k)$ of the form
\be
\rho^u_N(k) =   \rho(1)^k \left\langle \binom{j_1}{k}\right\rangle + \dots.
\ee
for any $k$.  Here the $\dots$ terms are positive definite. The expectation value $ \left\langle \binom{j_1}{k}\right\rangle $ is a sum over elements of $G_N$.  Consider just the contribution of the identity element, 
which has ${\bf j} (1) =(N,0,\ldots 0)$,
 to this expectation value. 
This gives 
\be
 \left\langle \binom{j_1}{k}\right\rangle 
 \geq  \binom{N}{k}|G_N|^{-1}\ 
\ee
so that
\be
|G_N| \ge { \rho(1)^k \over \rho^u_N(k)}   \binom{N}{k}~.
\ee
In the large $N$ limit, $\rho^u_N(k)$ must approach a finite number independent of $N$.  This implies that 
$|G_N|$ must grow faster than polynomially in $N$.  This demonstrates that the orbifold group $G_N$ must be rather large in order for the large $N$ limit to describe a well-defined theory of AdS gravity.
For example, this rules out the possibility that $G_N$ could be the cyclic group $\ZZ_N$ or even a product of a finite number of cyclic groups.

\subsection{Untwisted growth for permutation orbifolds}

We now wish to understand whether it is possible to constrain the density of states $\rho_N^u$ in the untwisted sector.
The untwisted sector is obtained by projecting out states in $\cH^{\otimes N}$ which that are not invariant under $G_N$. 
Since $G_N\subseteq S_N$, we therefore have
\be
\rho^u_{G_N}(\Delta) \ge \rho^u_{S_N} (\Delta)
~
\ee
for all $\Delta$. 
In this section we will find expressions for the number of states $ \rho^u_{S_N}$ in the $S_N$ untwisted sector (generalizing slightly the results of \cite{Dijkgraaf:1996xw}) and place a lower bound on the number of untwisted states in $G_N$.

For $S_N$, the coefficients $A_{\bf j}$ can be obtained using elementary combinatorics.  The result is
\be
Z^u_{S_N}(\tau) = \sum_{\bf j} {1\over (1^{j_1} j_1!) (2^{j_2} j_2!)\dots (N^{j_N} j_N!)} Z(\tau)^{j_1}\dots Z(N\tau)^{j_N}
~.
\ee
It is useful to combine these into the generating function 
$ \mathcal{Z}^u$:
\bea
\mathcal{Z}^u\equiv
\sum_{N} Z^u_{S_N}(\tau) p^N 
&=& 
\sum_{j_i}  \prod_{i=0}^\infty {p^{i j_i} Z(i\tau)^{j_i}\over i^{j_i} j_i!}
\cr
&=& \exp\left\{\sum_{i=1}^\infty{1\over i} Z(i\tau) p^i\right\} \,.
\eea
This is the grand canonical partition function, where instead of fixing $N$ we fix a chemical potential $\log p$ conjugate to $N$.
For simplicity let us set $\tau = i\beta/2\pi$, so that
\bea
\mathcal{Z}^u
&=& \exp\left\{ \sum_{i=1}^\infty \sum_{\Delta} {1\over i} \rho(\Delta) (p e^{-\beta m})^i \right\} 
\cr
&=& \prod_{\Delta} \left(1-p e^{-\beta \Delta}\right)^{-\rho(\Delta)}
\eea

We can use this to extract the large $N$ behaviour.  
The easiest way to this is to take $p\to 1$ in the grand canonical partition function, multiplying by $1-p$ to correctly account for the contribution of the ground state (following \cite{deBoer:1998us,Keller:2011xi}).  This gives
\bea
Z^u_{S_\infty}(\beta)&=&(1-p)\sum_{N} Z^u_{S_N} p^N\bigr|_{p=1}
\cr
&=& \prod_{\Delta>0} \left(1-e^{-\beta \Delta}\right)^{-\rho(\Delta)}\\
\cr
&=&
\exp\left\{
\sum_{i=1}^\infty {1\over i} (Z(i\beta)-1)
\right\}\ .
\eea
In the second line, the product over $\Delta>0$ is over all states other
than the vacuum.
We note that $Z(\beta)$ is monotonically decreasing
with $\beta$, and that $Z(i\beta)-1$ vanishes exponential at large $i$. 
From this it follows immediately
that $Z^u_\infty$ is finite for any finite $\beta$.  Thus there
is no Hagedorn divergence in the untwisted sector of $S_N$.  This is consistent with the long string picture, where the stringy states come from twisted sectors. 

It is convenient to rewrite this as an expression for the free energy of the untwisted sector in terms of the partition function of the original seed theory:
\be
F^u_\infty(\beta) \equiv \log Z^u_{S_\infty}(\beta) =  \sum_{i=1}^\infty {1\over i} (Z(i \beta )-1)
~.
\ee 
For small $\beta$
we can immediately estimate
\be
F^u_{S_\infty}(\beta) \sim Z(\beta) \sim \exp\left[\frac{c\pi^2}{3\beta}\right]\ .
\ee
We can then extract the density of states by evaluating the inverse Laplace transform
\be
\rho^u_\infty(\Delta-1) = \oint d\beta e^{\beta \Delta} e^{F^u_{S_\infty}(\beta)}\ ,
\ee 
where we integrate over the contour that goes through the saddle point given by the solution of the equation
\be
\Delta - \frac{c\pi^2}{3\beta^2}e^{c\pi^2/3\beta}=0\ .
\ee
The solution of $y = x^2 e^x$ is $x = \log y - 2\log\log y +\ldots$, so
\be
\frac{c\pi^2}{3\beta} = \log \frac{c\pi^2}{3} \Delta - 2\log\log\frac{c\pi^2}{3} \Delta+\ldots
\ee
at the saddle point.  This gives the saddle point estimate for the density of states
\be \label{SNuntwisted}
\rho^u_\infty(\Delta) \sim \exp \left(\frac{c\pi^2 \Delta /3}{\log c\pi^2 \Delta /3} \right)
\ee
for $S_N$.  
This saddle point approximation will be good when the argument of the exponential is large, i.e. when $c \Delta$ is large.

Equation (\ref{SNuntwisted}) is a lower bound for the number of states in the untwisted sector for any $G_N\subseteq S_N$.  Comparing to (\ref{HigherD}),
we conclude that, for any $G_N$, the density of states grows more quickly than is allowed in any local QFT in $d$ dimensions.
We note that the growth (\ref{SNuntwisted}) is still sub-Hagedorn.  In order to obtain a genuine Hagedorn spectrum we must consider  twisted sector states.

\section{The twisted sector partition function}

The partition function built out of only the untwisted sector states is not modular invariant.
Thus, in any orbifold CFT, twisted sector states must be introduced to
restore modular invariance. 
We will now study these twisted states. 
In this section, we will find it convenient to write the torus partition function using  
cylinder normalization rather than plane normalization:
\be
\tilde{Z}(q,\bar{q}) = \sum_{h,\bar{h}} \rho(h,\bar h) q^{h-\frac{c}{24}} \bar{q}^{\bar h-\frac{\bar{c}}{24}}
=\sum_{\tilde{h},\tilde{\bar{h}}} \tilde{\rho}(\tilde{h},\tilde{\bar h}) q^{\tilde{h}} \bar{q}^{{\tilde{ \bar {h}}}}
\ee
Here $\tilde{h},\tilde{\bar{h}}$ are the shifted dimensions, so that the vacuum state has  $\tilde{h} = \tilde{\bar{h}} = -c/24$.  With this normalization the  partition function $ \tilde{Z}$ will be invariant under $SL(2,\ZZ)$ modular transformations of the torus.

The full partition function, including twisted sector states, takes the form  \cite{Ginsparg:1988ui}
\be\label{orb}
\tilde{Z}_{G_N}=\sum_{[g]} \frac{1}{|C_g|}\sum_{h\in C_g}\begin{array}{r}\\  h~  \begin{array}{|c|}\hline ~ \\ \hline \end{array} \\  g~ \end{array}
\ee
where the summand is the torus partition function with insertions of $h$ along the (Euclidean) time cycle and $g$ along the spatial cycle. 
In this expression the twisted sector states are labelled by conjugacy classes  $[g]$ of $G_N$, which are inserted in the spatial direction.  When $g=1$ this is just the contribution from untwisted states computed in section~2. 
To project on to $G_N$ invariant
states in the $g$ twisted sector, we have summed over all $h$ in the centralizer
$C_g$ (i.e. over all $h$ that commute with $g$) inserted in the Euclidean time direction.
Under a modular transformation, each term in the sum will transform as
\be
\gamma  :
\begin{array}{r}\\  h~  \begin{array}{|c|}\hline ~ \\ \hline \end{array} \\  g~ \end{array}
\rightarrow \begin{array}{r}\\  h^a g^b~  \begin{array}{|c|}\hline ~ \\ \hline \end{array}~ \\  h^c g^d \end{array}\ 
\qquad \rm{for}\quad \gamma = \left(\begin{array}{cc}a&b\\c&d\end{array}\right) \in SL(2,\mathbb{Z})
\ ,
\ee
as long as $h$ and $g$ commute. From this one can  check that the result (\ref{orb}) is modular invariant.

\subsection{Symmetric Orbifolds: Hagedorn behavior from twisted states}

As a warmup, let us begin by considering the twisted states of the cyclic group $\mathbb{Z}_N$ and the symmetric group $S_N$. Let us first consider the case of  $\ZZ_N$, where each element is in its own conjugacy class. If for simplicity we take $N$ prime, then it is reasonably easy to compute (\ref{orb}) directly.  The result is  \cite{Klemm:1990df}:
\be
\tilde{Z}_{\ZZ_N}(\tau) = {1\over N} \tilde{Z}(\tau)^N + {N-1\over N} T_N \tilde{Z}(\tau) \ 
\ee
where we have defined the Hecke operator
\be\label{Hecke}
T_N \tilde{Z}(\tau) =  \sum_{d|N} \sum_{b=0}^{d-1} \tilde{Z} \left({N \tau + bd \over d^2}\right)\ .
\ee
The case of $N$ not prime is similar, but the expressions
are slightly more complicated.

For the case $G_N = S_N$, we can work out the
contribution of the twisted sectors to the 
full partition function explicitly.
It turns
to be easier to compute the grand canonical
ensemble partition function, \ie the generating function $\tilde{\mathcal{Z}}=\sum_N \tilde{Z}_N p^P$.
The result was found in 
\cite{Dijkgraaf:1996xw,Bantay:2000eq}:
\be \label{genF}
\tilde{\mathcal{Z} }= \sum_{N\geq 0} p^N \tilde{Z}_{S_N}(\tau) = 
\exp \left( \sum_{L>0} \frac{p^L}{L} T_L \tilde{Z}(\tau) \right) \ .
\ee
The interpretation is again that $T_L \tilde{Z}$ is the contribution
to the partition function of the twisted states coming from a cycle of length $L$.
The exponential arises because any element of $S_N$ can be viewed as a product of disconnected cycles.  Picking out the term $p^N$ from the exponential requires us to partition the integer $N$, giving the sum over all conjugacy classes $\bf{j}$.

One can then show that the twisted sector $({\bf 1})^{N-L}({\bf L})$ makes a contribution
\cite{Keller:2011xi}  
\be\label{Lcontribution}
\rho(\Delta) \sim e^{2\pi \Delta}
\ee
to the density
of states  at $\Delta = \frac{cL}6$.  This is a Hagedorn spectrum.  
The twisted states coming from elements of the form $({\bf 1})^{N-L}({\bf L})$, where $L$ is large, are known as long strings.

\subsection{General Case: Long Cycles and Hagedorn growth}

We will now see that this behaviour -- the existence of long cycles which lead to Hagedorn growth -- holds for a generic permutation orbifold $G_N$.

For a permutation group $G_N$, the orbifold partition function (\ref{orb})
can be written slightly more explicitly as \cite{Bantay:1997ek}:
\be\label{Bantay}
\tilde{Z}_{G_N}(\tau) = \frac{1}{|G_N|}\sum_{hg=gh} \prod_{\xi\in O(g,h)}\tilde{Z}(\tau_\xi)
\ee
The sum here is over all $g,h\in G_N$ which commute; as in (\ref{orb}), we can think of the sum over $g$ as labelling the twisted sector states and the sum over $h$ as projecting onto the $G_N$ invariant states in a given twisted sector.
A pair of commuting elements $g,h$ generate an Abelian subgroup of $S_N$,
which acts on the set $\{1,\ldots,N\}$ by permutation
of the elements.
In equation (\ref{Bantay}) we have denoted by $O(g,h)$ the set of orbits of this action. For each orbit $\xi \in O(g,h)$
we define the modified modulus $\tau_\xi$ as follows.
First, let $\lambda_\xi$ be the size of the $g$
orbit in $\xi$, and $\mu_\xi$ the number of $g$ orbits in $\xi$,
so that $\lambda_\xi \mu_\xi =|\xi|$.
Let $\kappa_\xi$ to be the smallest non-negative integer
such that $h^{\mu_\xi} g^{-\kappa_\xi}$ is in
the stabilizer of $\xi$.
Then 
\be
\tau_\xi = \frac{\mu_\xi\tau + \kappa_\xi}{\lambda_\xi}\ .
\ee
The product of the $\tilde{Z}(\tau_\xi)$ appearing in (\ref{Bantay}) is a generalization of the product of the $\tilde Z\left({N\tau + bd \over d^2}\right)$ which would appear when you exponentiate the Hecke operator (\ref{Hecke}) in (\ref{genF}).
It is straightforward to check that the terms in (\ref{Bantay}) with 
$g=1$ reproduce the untwisted sector partition function
(\ref{Polya}), since an element $h$ which has cycle
decomposition $\bf{j}$ has $j_i$ orbits $\xi_i$ of length $i$,
with $\kappa_{\xi_i}=0, \lambda_{\xi_i}=1$
and $\mu_{\xi_i}=i$.  So in this case $\tau_{\xi_i}= i\tau$.

We will now show that a cycle of length $L$ will lead to a Hagedorn density
at $\Delta = cL/6$.  We will set $c=\bar{c}$ for simplicity, although that does not effect our overall result.
 To begin, let us assume that
there is again an element $g$ of the form $({\bf 1})^{N-L}({\bf L})$
in $G_N$. The term with $(h=1,g)$ in (\ref{Bantay})
gives a contribution to the partition function
\be \label{GNdensity}
\tilde{Z}(\tau)^{N-L} \tilde{Z}(\tau/L) = e^{\beta c (N-L)/12}\tilde{Z}(\tau/L) + \ldots ,
\ee
From the first term we get a contribution 
\be
\rho_{N}(\Delta) = \tilde{\rho}(L(\Delta-cL/12))
\ee
to the density of states at weight $\Delta$. For $\Delta = cL/6$ and
$L$ large we are in the Cardy regime of $\tilde{Z}$, namely
\be
\tilde\rho(\tD)\sim e^{2\pi \sqrt{\frac{c \tD}{3}}} \ \ \ \ \ \ \ \ \ \tD\gg c\ .
\ee
Using \rref{GNdensity} we get
\be
\label{hag}
\rho_{N}(\Delta) \sim e^{2\pi cL/6} = e^{2\pi \Delta}\ ,
\ee
just as in the symmetric orbifold case.

We conclude that, if there is an element of the form $({\bf 1})^{N-L}({\bf L})$, then there will be a Hagedorn density at $\Delta = cL/6$.
These elements therefore constitute ``long strings," just as in the symmetric product case.
If $G_N$ has
elements of the form $({\bf 1})^{N-L}({\bf L})$ for all values of $L$ distributed  densely over the range from 1 to $N$ then there will be Hagedorn behaviour starting at energies of order the AdS scale.  
Indeed, in section~2.2 we saw that the group $G_N$ must have cycles of arbitrarily long length as $N\to\infty$.
This followed from the finiteness of $\rho_\infty$, which was necessary in order to have a well-defined large $N$ limit.  We see that the existence of this limit implies that $G_N$ is so large that twisted sector states lead to Hagedorn behaviour.  Thus, in this sense, every family of permutation orbifolds with a large N limit is a string theory.

There is one small caveat to the above argument: 
in section~2.2 we proved that there were cycles of arbitrary length, but we did not prove that they are necessarily of the form $({\bf 1})^{N-L}({\bf L})$. As we will show now, even if the elements are not of this form there will still be Hagedorn behaviour (\ref{hag}) with Hagedorn temperature equal to the AdS scale.
However, it might be that this Hagedorn behaviour would not dominate the spectrum until energies are Planckian $E\sim M_{Pl}$.   Thus the Hagedorn temperature might not be visible in $\rho_\infty$. We have not found any explicit examples where this occurs -- in all cases we have studied the Hagedorn behaviour (\ref{hag}) occurs even at the AdS scale.  But we cannot rule out the possibility that the Hagedorn regime occurs only for states above some other energy scale $E_1$ which is large in AdS units.

To see this, consider a cycle of the form $g=({\bf L}_1)\ldots ({\bf L}_I)$ with % which
\be \label{cycsum}
\sum_{i=1}^I L_i = N
\ .\ee 
The contribution from the $g$ twisted sector is
\be\label{Zprod}
\prod_{i=1}^I \tilde{Z}(\tau/L_i)
~.\ee
We want to compute the contribution to
a state with total weight $\tD >0$. The full density of states will be the convolution of the densities of the various factors appearing in (\ref{Zprod}). We will choose our $g$ such that
the majority
of the contributions in (\ref{cycsum})
come from long cycles with $L_i \gg 1$, so that we can use the Cardy formula to estimate (\ref{Zprod}). 
Then by the usual convolution arguments\footnote{A quick way to derive this formula is to note that we are effectively computing
the Cardy behavior of the tensor product of $I$ theories
with central charges $c_i = cL_i$, which has central charge $c=\sum_i c_i$.}
\be
\tilde\rho_N(\tD) \sim \exp2\pi\sqrt{c\left(\sum_i L_i\right)\tD/3} =
\exp(2\pi\sqrt{cN\tD/3})\ .
\ee
It follows that for $\tD = cN/12$ we reproduce the Hagedorn behaviour 
\be
\tilde\rho_N(\tD) \sim \exp 2\pi cN/6 = \exp 2\pi \Delta\ .
\ee
This establishes that for any permutation
orbifold there is a regime of Hagedorn behavior
which comes from long string states.

Unfortunately this does not immediately imply that the Hagedorn growth starts at the AdS scale, rather than at some other energy scale $E_1$. 
In particular, in the argument around (\ref{cyclecondition}) 
we have not 
shown that $m_\infty(i) > 0$ for most $i$.
We believe that it should be possible to prove this.
In particular, we believe that it should be possible to
strengthen the
bound (\ref{nonlocal}) to show
that any permutation orbifold has
Hagedorn density already at finite $\Delta$
even in the $N\to\infty$ limit.

\bigskip

\textbf{Acknowledgments:} We thank N. Arkani-Hamed, T. Banks, P. De Lange, M. Gaberdiel, L. Rastelli, E. Verlinde and H. Verlinde for useful discussions.  A. B. and A. M. are especially grateful to J. Lapan and A. Lepage-Jutier for collaboration at an early stage of this project.  C. K. is supported by the Rutgers
New High Energy Theory Center and by U.S. DOE Grants No.~DOE-SC0010008,
DOE-ARRA-SC0003883 and DOE-DE-SC0007897. A. B.
is supported by the Swiss National Science Foundation.
A. M. is supported by the National Science and Engineering Council of Canada.

\bibliographystyle{ytphys}
\bibliography{ref}

\end{document}